# Flow Dynamics of the Transversely Oscillating Tapered Circular Cylinder under Vortex-Induced Vibrations at low Reynolds number

Mayank Verma[1,2], Ashoke De[2,3 a)]

[1]*Current affiliation: IIHR – Hydroscience and Engineering, University of Iowa, Iowa City, Iowa 52242, USA*

[2]*Department of Aerospace Engineering, Indian Institute of Technology Kanpur, 208016, Kanpur, India.*

[3]*Department of Sustainable Energy Engineering, Indian Institute of Technology Kanpur, 208016, Kanpur, India.*

a) *Corresponding Author: ashoke@iitk.ac.in*

This study numerically investigates the influence of the taper on the flow-induced vibrations of an elastically mounted circular cylinder under Vortex-induced vibrations (VIV). The dynamic response of three different taper ratios ($TR$ defined as $TR = l/(d_2 - d_1) = 12$ (highly tapered cylinder), 20 (medium tapered cylinder), and 40 (low tapered cylinder)), where $d_1$ and $d_2$ are the largest and smallest diameter of the tapered cylinder with length '$l$' is studied at a fixed Reynolds number ($Re_D$, defined based on the averaged cylinder diameter, $D = (d_1 + d_2)/2$) of 150. The amplitude and frequency response of the tapered cylinder is characterized by a low mass ratio (defined as the ratio of the total oscillating mass to the displaced fluid mass) of m* = 2 over the wide range of reduced velocity ($3 \leq U_r \leq 16$) covering the full amplitude-response spectrum (based on the oscillation amplitude) of the VIV. The results show the existence of difference in the spanwise shedding of vortices owing to the poor spanwise pressure correlation. The flow field analysis in the wake of the oscillating cylinder reveals the dominance of the three-dimensional structures in the wake (near the top end with the larger diameter) behind the cylinder with the increase in the taper ratio (even at such low $Re_D$ where the uniform cylinder exhibits the two-dimensional wake). Also, the tapered cylinder exhibits a wide range of frequency synchronization (i.e. wide lock-in area) compared to the uniform cylinder. Tapering the cylinder results in the shift of the peak of the max oscillation amplitude or, in turn, the shift in the transitioning of the response branches. Further, force decomposition, energy transfer, and phase dynamics are also discussed for the taper cylinders.

## I. INTRODUCTION

Vortex shedding behind the bluff bodies has attracted many researchers due to its rich fluid physics and wide engineering application. Until the collapse of the Tacoma Bridge some fifty years ago, the vast majority of civil engineers were utterly unaware of the vortex-shedding phenomena behind bluff bodies and its potentially dire consequences. If the bluff body is elastically mounted, these shed vortices may cause structural oscillations, termed vortex-induced vibrations. If the frequency of the structural oscillations is close to the vortex shedding, high amplitude oscillations are observed, and this phenomenon is cited as 'lock-in' in the literature[1-3]. The vortex shedding behind the bluff body may be two-dimensional or three-dimensional, depending on the flow parameters/geometry. Williamson's study[4] reports the two-dimensional wake behind the stationary uniform circular cylinder for the flows with Reynolds number (Re) less than 178. The wake transitions to three-dimensional as Re exceeds 178 for the stationary cylinder. For the elastically mounted circular cylinder undergoing 1-DOF VIV[5,6], this transition from the two-dimensional wake to the three-dimensional wake takes place around Re = 300, while for 2-DOF VIV[7], it occurs around Re = 250. A three-dimensional wake may also occur behind the circular cylinder at much lower Re values if there is a spanwise variation in the incoming flow velocity (the uniform circular cylinder in sheared flow)[8] or the spanwise variation in the cylinder diameter (the tapered cylinder placed in the uniform flow)[9,10]. The case of sheared incoming flow is characterized by a shear parameter ($\beta = \frac{D}{u_m}\left(\frac{\partial u}{\partial z}\right)$ where D is the cylinder diameter, $u_m$ is the average incoming flow velocity, and z is the spanwise direction). At the same time, the taper ratio ($TR = l/(d_2 - d_1)$, where l is the length of the cylinder, $d_2$, and $d_1$ are the largest and smallest diameter of the tapered cylinder) is used to characterize the tapered cylinder. The case of the sheared flow differs from the uniform flow over the tapered cylinder because of the pressure-driven secondary spanwise flow in the sheared flow case[11]. The circular cylinder placed in the spanwise sheared incoming flow exhibits the spanwise cellular wake with the variation in the shedding frequency at the boundary of the two cells[8,12]. The cellular wake is also witnessed behind the linearly tapered circular cylinders[9,10]. Many engineering structures, such as offshore lighthouses, towers of wind turbines, industrial chimneys, etc., use tapered cylinders. Despite being geometrically simple, it produces complex flow patterns due to the spanwise variation in the local Reynolds and Strouhal numbers.

The pioneering experimental study of Gaster[9,10] on stationary tapered cylinders with a taper ratio of 36:1 and 18:1 presented the modulated velocity fluctuations in the cylinder wake and proposed the vortex cells (a region of constant vortex shedding frequency)[13,14] along the tapered cylinder. The spanwise variation in the vortex shedding frequency determines the size of these vortex cells: rapid frequency shift indicated cell change[15]. The experimental study of Hsiao and Chiang[13] indicates the Strouhal number (based on the local diameter of the tapered cylinder)-taper ratio dependence. They performed experiments on four different taper ratios (varying from vary from 24:1 to 75:1 at $4\times10^3 \leq \text{Re}_D \leq 1.4\times10^4$, where $\text{Re}_D$ is based on the average diameter ($D = (d_1 + d_2)/2$): $\text{Re}_D = U_\infty D/\upsilon$, $U_\infty$: the incoming flow velocity, and $\upsilon$: the kinematic viscosity of the fluid) and observed constant-frequency cells and transition intersection regions. Williamson[16] observed the oblique shedding behind uniform circular cylinders and reported that the end effects caused it. He further proposed that a discontinuity in the Strouhal-Reynolds number relationship results from transitioning from one oblique vortex-shedding mode to another.

Piccirillo and Van Atta[15] reported that the oblique vortex shedding in the tapered cylinder remains unaffected by the end conditions and has a geometric origin. They interlinked the bending of the vortex lines with the vortex splitting phenomena and reported it to cause a decrease in the local vortex shedding frequency. In a stationary tapered cylinder's wake and front stagnation areas, the spanwise pressure gradient drives spanwise secondary flows[17]. The numerical study by De & Sarkar[18] found random vortex splits and dislocations in the shedding signature of the stationary tapered cylinder and shows that pressure indirectly affects secondary instability over the stationary tapered cylinder.

Further, the wake behind the tapered cylinder shows different vortex-shedding patterns and three-dimensional structures when it oscillates. The wake visualization for the forced oscillations of the tapered cylinders (with a taper ratio of 40:1 at $400 \leq \text{Re}_D \leq 1500$) was reported by Techet et al.[19]. They obtained the dominance of a single frequency over the entire cylinder span in the lock-in region. They also reported the occurrence of combined 'hybrid' shedding with 2S (two single vortices per cycle of oscillation) type of shedding (around the larger diameters) and 2P (two pairs of vortices per cycle of oscillation) type of shedding (around the smaller diameters). Later, Hover et al.[20] continued to study the free-vibrations of the elastically mounted tapered cylinder (with a taper ratio of 40:1 at $\text{Re}_D \approx 3800$) and validated the results obtained for the forced harmonic oscillations for amplitude response and the phase between the lift force and the oscillations. They further reported that the tapered cylinders allow the lock-in at lower reduced velocities than the uniform cylinder case. Balasubramanian et al.[21] studied the VIV of tapered, pivoted cylinders in uniform and sheared flows and reported that structural properties primarily regulate the oscillation amplitudes and are insensitive to flow parameters. In their experimental study of VIV of the tapered cylinder (with a taper ratio of 20:1 at $1.4 \times 10^3 \leq \text{Re}_D \leq 7.02 \times 10^4$), Zeinoddini et al.[22] observed a wider lock-in range than its equivalent uniform cylinder. Seyed-Aghazadeh et al.[23] studied the cross-flow oscillations of the tapered cylinders (with taper ratios of 29:1, 17:1, and 10:1 at $370 \leq \text{Re}_D \leq 2300$) for the constant mass ratio of 7.5. They found that the maximum oscillation amplitude was independent of the taper ratio studied. Kaja et al.[24] performed the three-dimensional numerical simulations on the tapered cylinders (with a taper ratio of 20:1 at $\text{Re}_D = 500$) and observed the hybrid shedding with a broader lock-in regime. They further reported synchronizing the vortex shedding and sectional lift coefficient frequencies with the cylinder's vibration frequency during the lock-in regime.

The preceding investigations predominantly focused on stationary tapered circular cylinders, delving into the examination of wake flow dynamics and vortex shedding patterns surrounding the cylinders, including phenomena such as vortex dislocations, splitting, cellular vortex shedding, and oblique vortex shedding. However, scant attention has been devoted to exploring the phenomenon of vortex-induced vibration, specifically in tapered cylinders. It is crucial to study the flow separation and vortex shedding mechanisms around tapered cylinders (under VIV) and their impact on vibration characteristics. The research on flow-induced vibrations of tapered cylinders has made significant progress over the years, but several gaps still exist in understanding the fluid dynamics associated with this phenomenon. This study sets the stage for a comprehensive study to address the research gaps in the fluid dynamics of flow-induced vibrations of tapered cylinders. The author attempts to address the following relevant questions: (i) How does the flow separation, vortex shedding, and their interaction vary with the taper ratio? (ii) What is the effect of different taper ratios on the oscillation amplitude (transition of different amplitude branching) and frequency response of the cylinder (especially in the lock-in regime)? (iii) What role does the energy transfer between the cylinder and the fluid play in the oscillations? The authors try to address these questions by performing the numerical simulations on the 3D geometry of the tapered cylinders at $Re_D = 150$ over parametric variations such as taper ratios (TR) and reduced velocities ($U_r$) and assess their effects on the aerodynamic performance (in terms of the force coefficients, spanwise pressure distribution and its correlation, and the associated vortex shedding) and the VIV characteristics (in terms of the oscillation amplitude branching response, frequency (lock-in) response, and the phase dynamics associated).

## II. PROBLEM STATEMENT

The present article investigates the vortex-induced vibration characteristics of a tapered cylinder elastically mounted with the help of the spring-damper arrangement at a Reynolds number of 150. The spring-damper arrangement restricts the streamwise motion and allows only the cylinder's transverse oscillations, making the whole system one degree of freedom motion (1-DOF). The cylinder length (l) is 20*D, and the taper ratio ($TR = l/(d_2 - d_1)$) varies. The non-dimensional mass ratio ($m^*$, defined as the ratio of the structural mass to the mass of the displaced fluid) is 2; the damping ratio is taken as 0. The flow equations are discretized using a finite volume approach and solved over a three-dimensional computational domain. The typical computational domain and the models of the taper cylinders used for the present study is depicted in Fig. 1 (a). It spans between $-20 \leq x/D \leq 30$ in the streamwise direction, $-20 \leq y/D \leq 20$ in the transverse direction, and $0 \leq z/D \leq 20$ in the spanwise direction. The cylinder is located at the origin at t = 0. The incoming flow is assumed to be uniform and steady with a constant velocity. The slip condition is applied to the spanwise and transverse boundaries to reduce the effect of the computational domain boundaries, which is popularly used in the literature[25,26]. The outlet is treated with the advective boundary condition, a non-reflective boundary condition for velocity that solves the Euler equation $\frac{d\phi}{dt} + U\frac{d\phi}{d\eta} = 0$ at the exit boundary, where '$\eta$' is the outward-pointing unit normal vector. The 'MovingWallVelocity' boundary condition is applied to the cylinder wall to incorporate the motion of the cylinder.

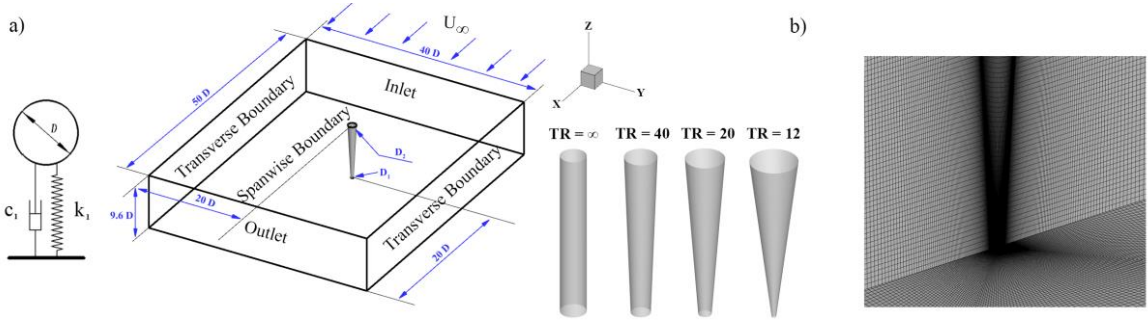

Fig. 1. (a) Schematics of computational domain with elastically mounted tapered cylinder (b) zoomed view of computational mesh used for the present study

## III. NUMERICAL DETAILS

### A. Governing Equations

The incompressible form of Navier-Stokes equations is solved assuming the flow to be incompressible, viscous, and laminar over a three-dimensional computational mesh (shown in Fig. 1 (b)) using a finite volume-based open-source solver of OpenFOAM. The equations for the flow are:

$$\nabla \cdot \vec{v} = 0 \tag{1}$$

$$\rho_f \left[ \frac{\partial \vec{v}}{\partial t} + (\vec{v} \cdot \nabla)\vec{v} \right] = -\nabla p + \mu_f \nabla^2 \vec{v} \tag{2}$$

where $\vec{v}$ is the velocity vector, $\rho_f$ is the fluid density, $p$ is the static pressure, and $\mu_f$ is the dynamic viscosity of the fluid. The Reynolds number ($Re_D$) value is fixed at 150 for the present study. The pressure-velocity coupling is felicitated using the Pressure Implicit Method for Pressure Linked Equations (PIMPLE) algorithm. Second-order discretization schemes are invoked to address all the spatial and temporal terms in the governing equations.

To incorporate the rigid cylinder's motion, a second-order linear spring-mass-damper equation is solved:

$$m\frac{d^2Y}{dt^2} + 2m\zeta\omega_n\frac{dY}{dt} + kY = F_Y \qquad (3)$$

where Y represents the cylinder displacement in the transverse direction, m is the structural mass of the cylinder, $\zeta$ is the structural damping coefficient, $\omega_n$ is the natural frequency of the oscillating cylinder (can be written as $\omega_n = \sqrt{k/m}$) and k is the stiffness coefficient. $F_Y$ represents the total transverse force component (pressure + viscous forces). For the simplicity of the calculations, the authors have assumed the springs to be linear and the damping coefficient to be zero (to allow maximum oscillations). Some of the non-dimensional parameters are used to effectively describe the results: non-dimensional mass ratio ($m^* = m/\frac{\pi}{4}\rho_f D^2 L$), non-dimensional reduced velocity ($U_r = U_\infty/f_n D$), frequency ratio ($f^* = f_y/f_{ny}$), non-dimensional frequency ($f_n^* = f_n D/U_\infty$), non-dimensional time ($\tau = tU_\infty/D$), non-dimensional mean oscillation amplitude ($A_Y^*$), and the phase difference between the lift force and the oscillation amplitude ($\phi = \phi_{F_y} - \phi_y$), calculated from the mean envelope of the Hilbert transform of the amplitude response.

Eqs. 1-3 need to be solved in a coupled manner to calculate the system response accurately. The fluid force is obtained from the fluid solver in each time step, and the motion of the cylinder is calculated by the solution of Eq. 3 using the fourth-order Runge-Kutta scheme. Thus, the velocity, acceleration, and the cylinder's position depend on the fluid load and make the system non-linear. The chosen method in this study is a staggered or loosely coupled method in which the flow and structural equations are solved independently and sequentially, with coupling invoked by forces and boundary conditions in the OpenFOAM framework. The loosely coupled methods are extensively used in the literature[27-30]. The detailed implementation of the weakly coupled form of the structural equation can be found in the work by Jester and Kallinderis[31] and Carmo et al.[32]. The timestep is selected based on the timestep independence study and is kept sufficiently small for better coupling convergence. To incorporate the mesh motion due to the movement of the cylinder, after each computational time step, the positions of the finite volume cells are computed by Laplace's equation,

$$\nabla \cdot (\gamma_m \nabla z) = 0 \qquad (4)$$

where $\gamma_m$ is the diffusion coefficient, and $z$ is the mesh-cell center displacement field. The mesh motion is distributed using the inverse mesh diffusion model based on the inverse distance from the cylinder[33-35]. The diffusivity field is calculated using the quadratic relation on the inverse of the cell center distance ($l$) to the nearest boundary, $1/l^2$.

## B. Grid Independence Study and Error Analysis

Before continuing the main study, we performed the grid-independence study for the tapered cylinder (with TR = 12) at $Re_D = 150$. The mass ratio of the cylinder is 2.0, with a damping coefficient of 0 (to maximize the oscillation amplitude). The reduced velocity is $U_r = 6$. The O-grid mesh is generated around the tapered cylinder using the ICEM-CFD tool in the near wake region. The cylinder wall is resolved up to a y$^+$ value of 1 to resolve the wall stresses accurately with a cell expansion ratio of 1.02 in the small rectangular region around the cylinder. For computing cost-saving, quadrilateral cells with an expansion ratio of 1.2 occupy the rest of the computational region. The grid independence study uses four grids: Grid-1 (~ 1.5 million cells), Grid-2 (~ 3.0 million cells), Grid-3 (~ 6 million cells), and Grid-4 (~ 12 million cells). Figure 2 (a) compares the variation of the rms value of the lift coefficient ($C_{L_{rms}}$) for different mesh sizes. The differences between Grid-3 and Grid-4 reduce to less than 1% (Table I). Thus, Grid-3 (with 6 million cells) should suffice for this study.

**Table I.** Grid Independence study for the moving cylinder at $Re_D = 150$ [Values in bold show the selected grid]

| Grid | No. of cells | $C_{L_{rms}}$ | % Change in $C_{L_{rms}}$ value | $A_Y^*$ | % Change in $A_Y^*$ value |
|---|---|---|---|---|---|
| Grid-1 | 1,553,136 | 0.3347 | - | 0.5169 | - |
| Grid-2 | 3,047,980 | 0.3242 | 3.2387 % | 0.5152 | 1.688 % |
| **Grid-3** | **6,012,118** | **0.3230** | **0.3715 %** | **0.5140** | **0.2334 %** |
| Grid-4 | 11,966,190 | 0.3227 | 0.0929 % | 0.5139 | 0.0194 % |

Further, to develop more confidence in the chosen grid, we also examined Roache's Grid-Convergence Index (GCI), extensively utilized in the literature[36,37]. A detailed description of the GCI calculations can be found in the previous papers of the research group[7,26,30]. For conducting the GCI study, we have looked at the parameter $C_{L_{rms}}$ with the safety factor as 1.25[38]. The results are tabulated in Table II (for the definition of symbols used, please refer to Ref. 26). We observe that the value of GCI goes down with the consecutive grid refinement from Grid-2 to Grid-4. This confirms that the Grid-3 is nicely resolved. Also, comparing the slope of the L$_2$ norm with the theoretical slope of order 2 shows the order of accuracy to be 1.92.

**TABLE II.** Richardson error estimation and grid-convergence index for three sets of grids

|  | $r_{32}$ | $r_{43}$ | o | $\varepsilon_{32}$ ($\times 10^{-3}$) | $\varepsilon_{43}$ ($\times 10^{-3}$) | $E_2^{coarse}$ ($\times 10^{-2}$) | $E_1^{fine}$ ($\times 10^{-2}$) | $GCI^{coarse}$ (in %) | $GCI^{fine}$ (in %) |
|---|---|---|---|---|---|---|---|---|---|
| $C_{L_{rms}}$ ($Re_D = 150$) | 1.25 | 1.25 | 1.92 | 3.701 | 0.929 | 1.062 | 0.174 | 1.327 | 0.217 |

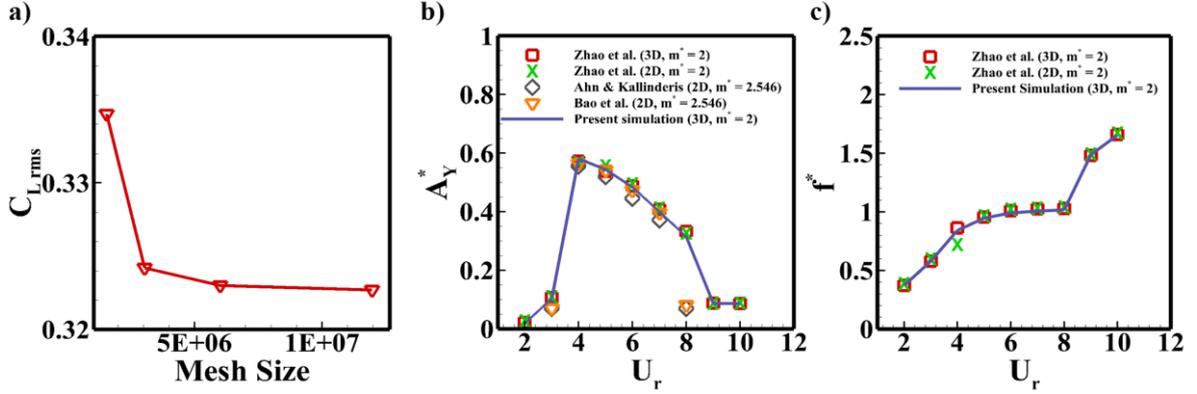

Fig. 2. (a) Grid Independence study based on the rms value of the lift coefficient for different mesh size at $U_r = 6$; Numerical validation study for a transversely oscillating circular cylinder: (b) oscillation amplitude, and (c) frequency response (Zhao et al.[6]; Ahn & Karllinderis[39]; Bao et al.[40]) [$m^* = 2, \zeta = 0, TR = 12, Re_D = 150$]

## C. Numerical Validation

To assess the accuracy and reliability of the numerical simulations being conducted within the framework of the study, the authors have duly validated the results of the flow-induced vibrations of the uniform cylinder (i.e. $TR = \infty$) with the available literature at low mass ratio and $Re_D = 150$. The oscillation amplitude response of the cylinder has been compared with the available data of Zhao et al.[6] [both two-dimensional (2D) and three-dimensional (3D) data at $m^* = 2$], Ahn & Kallinderis[39], and Bao et al.[40] [at $m^* = 2.546$] in Figure 2 (b). The results nicely capture the transition from the Initial branch (IB) to the Lower branch (LB) over the spectrum of reduced velocities and show an excellent agreement with the available literature. Further, the frequency response of the cylinder is compared with the available data of Zhao et al.[6] and also found to be in good agreement. The lock-in range for the uniform cylinder matches well with the available literature.

## IV. RESULTS AND DISCUSSION

The present study deals with the comprehensive study of the flow-induced motion of the tapered cylinder under 1-DOF VIV (in transverse direction) conditions. We have performed the three-dimensional CFD simulations at a Reynolds number of 150 to characterize the VIV characteristics at low Re. The reduced velocity ($U_r$) varies from 3 to 16, which captures the complete lock-in regime for all taper ratios. The structural damping coefficient ($\zeta$) is set to 0 (for maximum amplitude) with a mass ratio ($m^*$) value of 2.0 (corresponding to the low mass ratio value). The result and discussion part of the article is divided into two major sections where the first section deals with the fluid flow characteristics in the wake behind the oscillating tapered cylinders at a fixed reduced velocity of 6 using mean pressure distribution, the vorticity contours and the iso-surfaces of the vorticity. The second section deals with the VIV characteristics over a wide range of reduced velocities via means of the mean oscillation amplitude and frequency response, the force decomposition, and the phase dynamics of the tapered cylinders and compares it with that of the uniform cylinder case.

## A. Fluid Dynamic Response of Oscillating Tapered Cylinders

We have performed the three-dimensional numerical simulations for the three different taper ratios (TR = 12, 20, and 40) and have compared it with the uniform cylinder case ($TR = \infty$) at a fixed Re value (i.e. $Re_D = 150$) with a fixed reduced velocity ($U_r = 6$), which lies in the lock-in range. This section reports the tapered cylinder's fluid dynamics behavior and associated wake flow characteristics.

### 1. Oblique Vortex Shedding and Wake Three-dimensionality

The vortex shedding behind the uniform cylinder is well reported in the literature[16], where the cylinder sheds the alternate vortices/pairs of vortices over a cycle of oscillation. To visualize the flow structures behind the oscillating cylinder, we show the isometric view of the flow field (in the x-z plane) in the top row of Fig. 3 by plotting the iso-surfaces of the second eigenvalue[41] ($\lambda_2$) of the tensor $S^2 + \Omega^2$, where S is symmetric, and $\Omega$ is the antisymmetric part of the velocity gradient tensor, $\nabla u$. The eigenvalues are presented in the normalized form (concerning the flow velocity and the mean cylinder diameter, i.e. $e_2 = -\lambda_2 / (U / D_{mean})^2$). Further, the iso-surfaces are colored with the spanwise-vorticity ($-5 \leq \omega_z \leq +5$). For the case of low Re values (i.e. the present case of $Re_D = 150$), the shedding occurs uniformly over the whole span for the uniform cylinder case exhibiting a two-dimensional wake (see top row of Fig. 3 (a)). However, as we taper the cylinder, the incoming flow experiences the spanwise variation in the diameter, resulting in different advection velocities at different spanwise locations over the cylinder. The spanwise variation of the Reynolds number ($Re_z$) is observed to be in the following ranges based on the TR value: (i) for TR = 40: $112.5 \leq Re_z \leq 187.5$, (ii) for TR = 20: $75 \leq Re_z \leq 225$, and (iii) TR = 12: $25 \leq Re_z \leq 275$. For the case of the mild taper (i.e., TR = 40), as the $Re_z$ remains low and under the transition Re (at which the wake transitions from two-dimensional wake to three-dimensional wake) for the uniform cylinder, the wake is two-dimensional. But, due to the difference in the advection velocity spanwise, the vortex shedding becomes oblique (as can be seen in the case of TR = 40 in Fig. 3 (b)). The flow behind the transversely oscillating uniform circular cylinder is known to show three-dimensionality in the wake around the Re value of 200. As we increase the taper (from mild to intermediate or high), the Re range for the TR = 20 and 12 crosses the transition Re (i.e. $Re_z > 200$), and the three-dimensional structures (with cellular vortex patterns[9,10]) start to appear near the top end of the cylinders for these cases (Fig. 3 (c) and (d)) and wake behind the cylinder exhibit the three-dimensional nature. The three-dimensionality increases with the increase in taper. This alters the Strouhal number spanwise and suppresses/delays the vortex shedding at the bottom end of the cylinder (where $Re_z$ is relatively low).

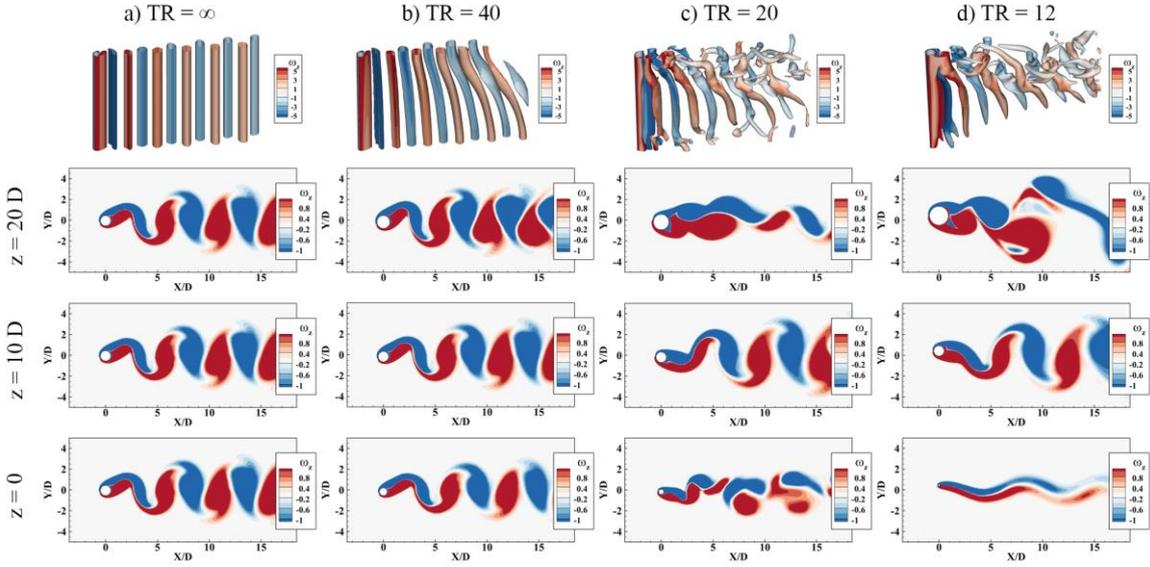

**FIG. 3.** Three-dimensional structures behind the transversely oscillating tapered circular cylinder for various taper ratios via means of the iso-surfaces of eigenvalue $e_2 = 0.01$ (top row); distribution of the instantaneous vorticity in the z plane at three different spanwise locations: bottom end (z/D = 0), midplane (z/D = 10), and top end (z/D = 20) for: (a) TR = $\infty$, (b) TR = 40, (c) TR = 20, and d) TR = 12. [$m^* = 2.0, Re_D = 150, U_r = 6.0, \zeta = 0, L = 20D$]

In addition to the iso-surfaces, Fig. 3 also depicts the instantaneous spanwise vorticity distribution behind the cylinder at three different spanwise locations: z = 0 (corresponds to the bottom end), z = 10 D (corresponds to the midplane) and z = 20 D (corresponds to the top end). We observe that due to the two-dimensional nature of the wake, the cylinders with TR = $\infty$ and 40 shed similar patterns of the alternate vortices over the whole span (with varying separation distances spanwise between the vortices of opposite sign for TR = 40), which resembles the 2S type vortex shedding (where a single vortex is shed each half-cycle of oscillation). As the three-dimensionality introduces the range of flow structures spanwise, the instantaneous vorticity distribution is different at different spanwise locations for TR = 20 and 12. For the case of TR = 20, the top end follows the 2S pattern of the vortex shedding (vortices being shed much closer to the cylinder), while the midplane resembles the standard Kármán Street kind of 2S pattern (for all TR values). For the case of the highly tapered cylinder (i.e. TR = 12), the vortex shedding is immensely suppressed at the bottom end of the cylinder due to the very low value of $Re_z$, while the top end sheds the 2S type vortex-shedding pattern. So, this highlights the appearance of different shedding present spanwise over the cylinder length. To further understand the vortex dynamics, we have compared the characteristics of the oscillating taper cylinder with the case when the cylinder is rigidly mounted (i.e., the case of the stationary taper cylinder).

### 2. Stationary vs VIV Motion of a Taper Cylinder

The preceding subsection shows the three-dimensional structures in the wake behind the tapered cylinder with high taper oscillating under 1-DOF VIV. In contrast, the uniform cylinder exhibits a purely two-dimensional wake at that value of Re. Fundamentally, two combined effects are working there, which might contribute to the three-dimensional nature of the wake: a) tapering the cylinder, which induces the geometric shear to the wake flow, and b) the oscillating nature of the cylinder, which imparts additional vortex and shear layer dynamics. Thus, to elucidate the role of the oscillating nature, we compared the wake behind the oscillating tapered cylinder (with TR = 12 at $U_r = 6$) with the stationary tapered cylinder in Fig. 4.

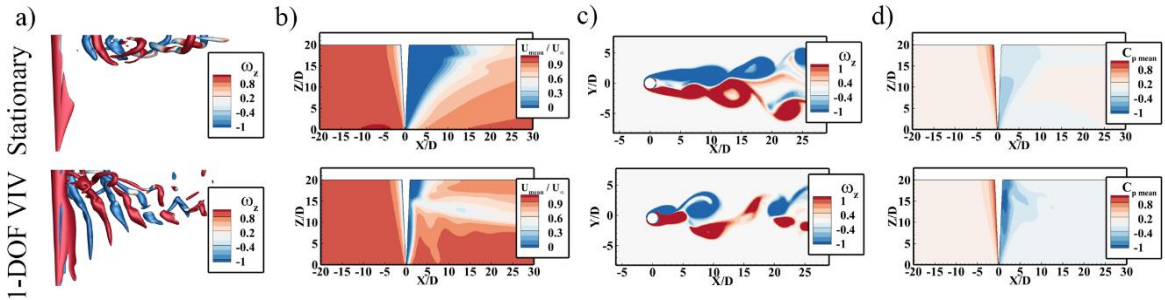

**FIG. 4.** Comparison between the characteristics of the taper cylinder under stationary and 1-DOF VIV motion (at $U_r = 6$): (a) Normalized second eigenvalue ($e_2$), (b) Time-averaged streamwise velocity contours ($U_{mean}/U_\infty$, in x-z plane), (c) Spanwise vorticity ($\omega_z$ in x-y plane), and (d) Time-averaged pressure coefficient ($C_{Pmean}$)

To depict the structures in the wake, we have plotted the normalized second eigenvalue ($e_2$) colored with the spanwise vorticity in Fig. 4(a). In the case of the stationary cylinder, the flow structures are present only near the top end (with a larger diameter where $Re_z \approx 275$). In contrast, for the rest of the cylinder span (approximately 80%), the shear layer with the positive vorticity seems to be attached. It exhibits the loss of any vortex shedding (due to a very low $Re_z$ ($\approx 25$) value near the lower end of the cylinder). This is not true for the case of the oscillating tapered cylinder. For this case, near the lower end, the vortices are two-dimensional and exhibit the obliqueness due to the spanwise difference in the advection velocity. At the same time, the top end shows the presence of similar three-dimensional structures as present in the stationary case and exhibits the vortex dislocations at the intersection point (where the two-dimensional wake becomes three-dimensional).

Further, Fig. 4 (b), (c), and (d) shows the distribution of the time-averaged streamwise velocity (in x-z plane), spanwise vorticity (near the top end, i.e., z/D = 20, in x-y plane), and the time-averaged pressure coefficient in the wake. We observe the presence of the negative values of the streamwise velocities in the near wake spanwise, which extends to the far wake in the case of the stationary cylinder. It shows that the shear layer remains attached till far downstream in the case of the stationary cylinder compared to the oscillating cylinder. Thus, to confirm this, we have plotted the spanwise vorticity near the top end of the cylinder. We observe that the oscillating cylinder, due to additional energy of the motion, sheds the vortices much closer to the cylinder (approximately at x/D = 5), while the breaking up of the vortices from the attached shear layer occurs much far downstream (approximately at x/D = 10). This causes the more energized near wake flow for the case of the oscillating cylinder and results in more negative pressure near the leeward surface of the cylinder (which can be seen in the contour of mean pressure in Fig. 4 (d)) over the whole span as compared to the case of the stationary cylinder. The temporal and spatial evolution of these shedding patterns is thoroughly discussed in the coming subsection to understand the development and shedding dynamics of the vortices.

## 3. Unsteady vortex dynamics and shedding modes

From the vorticity contours of Fig. 3, we can observe that the midplane for each TR value sheds vortices in the same pattern (i.e. 2S pattern) due to the equal mean diameter across all the taper ratios. By examining the time and the flow snapshots together, we may establish a connection between certain events in developing the vortices. Figure 5 reports the unsteady vortex dynamics at five distinct time instants (i.e. 0, T/4, T/2, 3T/4, and T; T is the period of oscillation) for different taper ratios at the top end (i.e. z/D = 20). Due to poor spanwise shear (owing to its geometry), TR = 40 sheds the vortices similar to the uniform cylinder (i.e. 2S pattern). As we increase the taper (i.e. TR = 20 and 12), the vortex shedding from the top end still follows the 2S pattern, but the shedding is much closer to the cylinder. For such cases, as the cylinder moves upwards (from t = 0), the vortex with negative vorticity starts to grow in size near the top surface of the cylinder. As the cylinder attains the maximum amplitude (at t = T/4), the vortex is no longer connected to the cylinder's surface and starts to advect in the flow. During the downward motion of the cylinder, there is the appearance of a small vortex with positive vorticity near the lower half of the cylinder as it passes through the origin (Y/D = 0 at t = T/2), which grows in size with the downward motion of the cylinder. As the cylinder attains the maximum amplitude (in the negative Y direction at t = 3T/4), the vortex with positive vorticity separates from the cylinder and advects into the flow. Thus, over each half-cycle, the cylinder's top end (for TR = 20 and 12) sheds one vortex resembling the 2S vortex shedding pattern. Since the shedding of the vortices at the top end for TR = 20 and 12 is very close to the cylinder's leeward surface, this affects the pressure distribution around the cylinder, which is discussed in detail in the coming subsection.

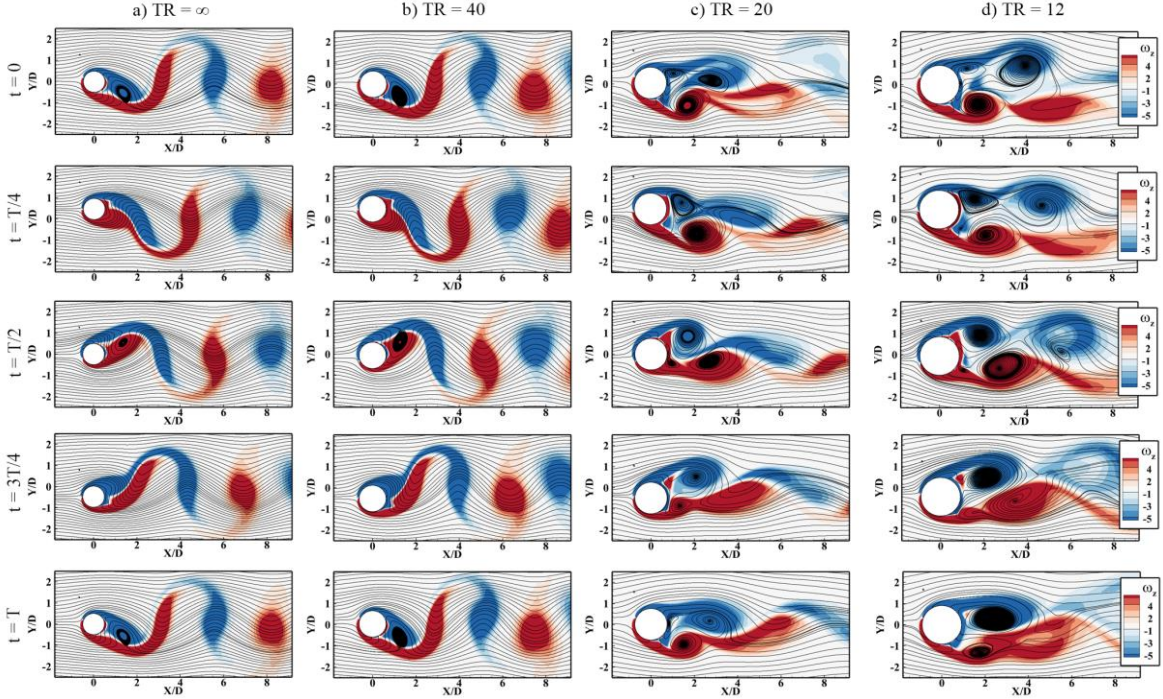

**FIG. 5.** Temporal evolution of the instantaneous spanwise vorticity ($\omega_z$) along with the velocity streamlines for one complete oscillation cycle at the top end (z/D = 20) of the cylinder with different taper ratios: (a) TR = $\infty$, (b) TR = 40, (c) TR = 20, and (d) TR = 12. [ $m^* = 2.0, \text{Re}_D = 150, U_r = 6.0, \zeta = 0, L = 20D$ ]

### 3. Spanwise distribution of Mean Pressure and Force Correlations

From the last two sub-sections, we have observed that the shedding varies spanwise as the spanwise shear becomes stronger for the tapered cylinders. This, in turn, affects the distribution of the mean pressure around the cylinder surface. Figures 6 (a), (b), and (c) compare the distribution of the mean pressure coefficient ($C_{PMean}$), which is calculated as $(\bar{p} - p_\infty)/(0.5\rho U_\infty^2)$, $\bar{p}$ representing the mean pressure at the specific location on the cylinder surface, and $p_\infty$ is the freestream pressure. It is calculated around the cylinder surface at three different spanwise locations: z/D = 0, 10, and 20, respectively, and the results are reported for the top side of the cylinder (i.e. $0^o \leq \theta \leq 180^o$) as for the low Re flows, the pressure distribution is symmetrical owing to the cylinder's circular geometry[43-45].

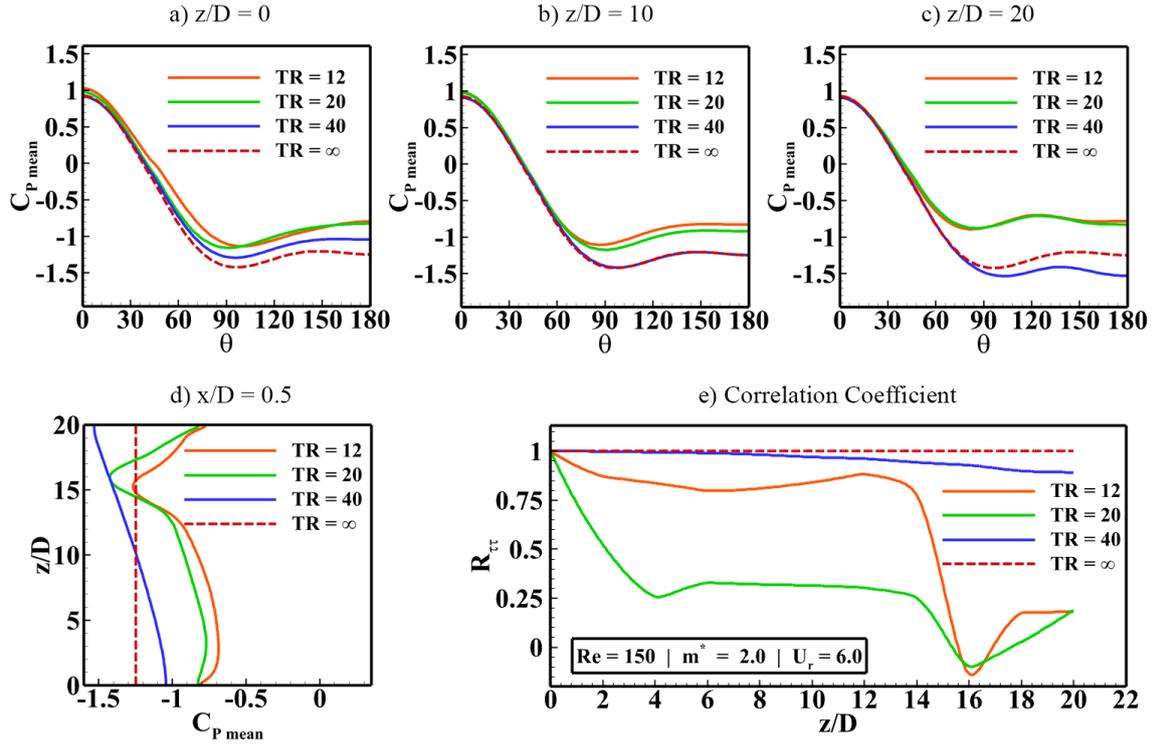

**FIG. 6.** Distribution of mean pressure coefficient ($C_{PMean}$) over the cylinder surface at three different locations: (a) z/D = 0, (b) z/D = 10, and (c) z/D = 20, (d) the spanwise distribution of the mean pressure coefficient over the leeward surface of the cylinders, and (e) spanwise correlation coefficient for different taper ratios [ $m^* = 2.0, \mathrm{Re}_D = 150, U_r = 6.0, \zeta = 0, L = 20D$ ]

The $C_{PMean}$ distribution over the cylinder surface $TR = \infty$ is consistent with the available literature. $C_{PMean}$ starts with the maximum value of 1.0 at $\theta = 0^o$, rapidly decreasing over the top surface (till $\theta \approx 90^o$), resulting in a favorable pressure gradient in the regime ($0^o \leq \theta \leq 90^o$). From the minimum $C_p$ value, it exhibits the regime of the adverse pressure gradient ($90^o \leq \theta \leq 135^o$) till the laminar boundary layer separation at $\theta \approx 135^o$ and the onset of the base region. The $C_{PMean}$ remains almost constant in the base region till the base point at $\theta = 180^o$. The tapered cylinder also follows a similar $C_{PMean}$ trend. The spanwise variation in the diameter of the cylinder results in the increase/decrease of the base pressure behind the cylinder at different spanwise locations, which is attributed to the vortex formation dynamics in the near wake and its shedding over the oscillation cycle (see Fig. 6 (d)). The pressure varies linearly over the span for the mild taper (i.e., TR = 40) due to the presence of the two-dimensional wake. For the moderate or high taper cases (i.e. TR = 20 or 12), three-dimensional structures alter the pressure profile's linear spanwise behavior. The dip in the pressure profile corresponds to the occurrence of the vortex dislocation in the wake behind the cylinder. Further, this spanwise pressure gradient drives the secondary flow spanwise, consistent with the available literature on the stationary taper cylinder[23,18].

To further understand the correlation between the spanwise pressure distribution and vortex dynamics, we have calculated the spanwise correlation coefficient based on the sectional pressure as follows:

$$R_{\tau\tau} = \frac{\text{Cov}(X,Y)}{\sqrt{\text{var}(X) \cdot \text{var}(Y)}} \qquad (5)$$

Where X and Y are the pressure signals at different spanwise positions, 'Cov' is the covariance between X and Y, and 'var' is the variance of the variable X or Y. The results for different cases are reported in Fig. 6 (e). For the uniform cylinder, the correlation coefficient remains constant at unity. As the taper ratio increases, the correlation coefficients decrease from one end of the cylinder to the other. For the case of the TR = 40, the correlation coefficient is almost constant till z/D = 10 and starts to decrease slowly afterward, indicating the oblique shedding with a two-dimensional nature of the wake. As the taper increases to TR = 20, 12, we identify three distinct regimes based on the spanwise correlation coefficient values. In the first regime, the correlation coefficient rapidly decreases spanwise till z/D = 4 and then becomes constant with a slight rise (for $6 \leq z/D \leq 14$). The close observation of the correlation coefficient with the flow snapshots of the iso-surfaces at z/D = 4 indicates that the vortices rearrange themselves into a standard von Karman vortex street type of shedding pattern and continue to shed vortices in that fashion till z/D = 16. At z/D =16, the variation in the spanwise correlation coefficient changes its sign (negative values) and indicates the presence of stronger secondary structures, which can be seen in the iso-surfaces of Fig. 4. The local Strouhal number ($St_z = f_z D / U_\infty$, where $f_z$ is the dominant frequency of the FFT of the transverse velocity at the given spanwise location) also shows a staircase profile along the span (not shown for conciseness) with the significant variation near the ends consistent with the available literature for the stationary tapered cylinders[17,18]. Further, the VIV characteristics of the tapered cylinder are discussed in the following section.

## 4. Oscillation amplitude and frequency response

Further, to access the VIV characteristics of the tapered cylinder, we have looked at the oscillation amplitude and frequency response for different taper ratios in Fig. 7. For the chosen VIV parameters ($m^*, U_r, \zeta$), we observe that the peak oscillation amplitude for all the taper ratios is almost identical, but the transverse force coefficients is quite different. This difference is attributed to the associated amplitude-response branches for the chosen parameters (i.e. TR = ∞ and 40 are in the lower branch (LB), and TR = 20, 12 are in the initial branch (IB)), which is discussed in detail in the future sections. To understand the different behaviors of the lift and oscillation amplitude, we have plotted the frequency response of the lift coefficient and oscillation amplitude. We observe that the cylinder oscillations with TR = ∞ and 40 show the single frequency while the lift coefficient shows the dominance of two peaks (first and third harmonics of the lift coefficient). As the taper increases, the contribution of the third harmonics diminishes, and the frequency response shows the dominance of a single frequency peak. Also, the frequency value reduces with increased taper, and the lock-in occurs at a frequency lower than 1 for the highly tapered cylinders.

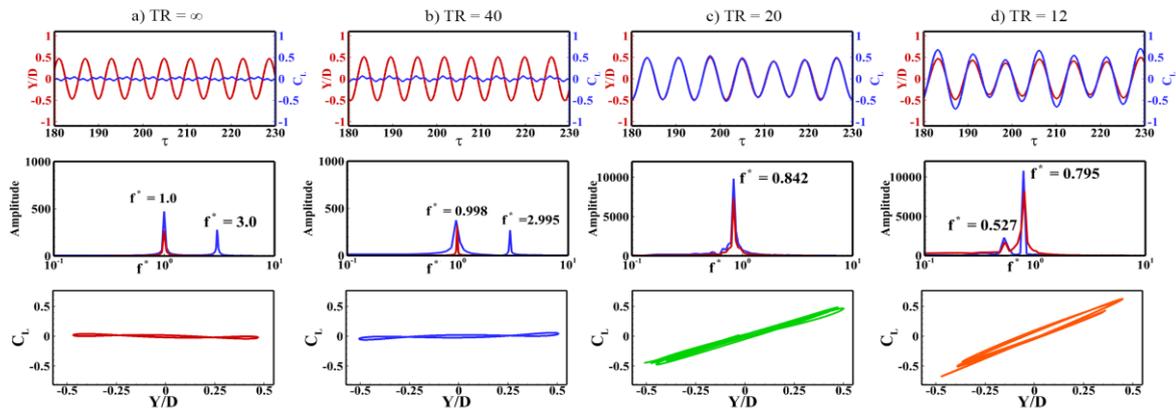

**FIG. 7.** Force, oscillation amplitude, and frequency response of the cylinder with different taper ratios: (a) TR = ∞, (b) TR = 40, (c) TR = 20, and (d) TR = 12. [ $m^* = 2.0, Re_D = 150, U_r = 6.0, \zeta = 0, L = 20D$ ]

Further, to understand the phase difference between the lift and the oscillation amplitude, we have also plotted the phase portraits (for five oscillation cycles) in the bottom row of Fig. 7 for different taper ratios. The phase portraits are used to understand the phase dynamics qualitatively. They can also be a good indicator of the nature of the wake (i.e. a two-dimensional or three-dimensional wake), especially for such low Re-value flows[21]. The cylinder with TR = ∞ and 40 sweeps the same path for all five oscillation cycles; this indicates that the structures the cylinder sheds into the wake while oscillating are two-dimensional and don't affect its oscillation path. The cylinder with TR = 20 and 12 doesn't follow the same trajectory for all five oscillation cycles, indicating the loss of two-dimensionality in the wake. The iso-surface plots in the first row of Fig. 3 confirm this visually. The presence of the three-dimensional structures in the wake results in the different forces acting on different spanwise cylinder locations, which is discussed in the coming subsection.

**5. Force Decomposition**

To further understand the force dynamics related to tapered cylinders, the temporal variation of the fluid force in the transverse direction (being responsible for the oscillation of the cylinder) and its associated force components is thoroughly investigated over one complete oscillation cycle. Figure 8 below depicts the distribution of the total cross-flow force ($\tilde{C}_y$) and its pressure ($\tilde{C}_y^P$) and viscous components ($\tilde{C}_y^V$) on the left axis (with green line) with the oscillation amplitude on the right axis over one complete cycle (where T is the period of one oscillation) at $U_r = 6$. There are some similar features between all the cases. In all cases, the cylinder's oscillations are sinusoidal. We observe a substantial difference in the variation of the total cross-flow force and its decomposed components for different taper ratios. The first visible difference is in the amplitude of the total cross-flow force and its components. The cylinder with TR = $\infty$ and 40 depicts a much lower amplitude of the cross-flow force than the highly tapered cases (i.e. TR = 20 and 12). For these cases, the motion of the cylinder is observed to be governed by the viscous component of the flow, the amplitude of the pressure component being of the order of the viscous component. Also, a phase lag of 180° exists between these components, reducing the total cross-flow force amplitude. For the cases of the highly tapered cylinders (TR = 20 and 12), the cylinder oscillation is majorly governed by the pressure component of the total cross-flow force (the viscous component's contribution is relatively low compared to the pressure component: see Fig. 8). Also, these components are almost in phase with each other (and with the oscillation amplitude), contributing to the enhanced total cross-flow force values at any time instant.

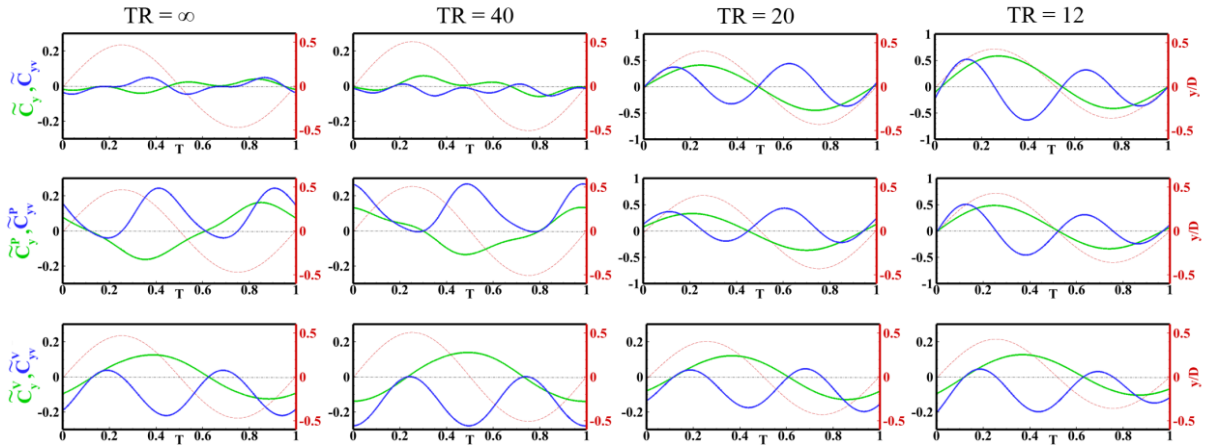

**FIG. 8.** Temporal evolution of the cross-flow force fluctuation (green line, left axis) and cylinder displacement fluctuation (red dotted line, right axis) for different tapper ratios: TR = $\infty$, TR = 40, TR = 20, and TR = 12. ($\tilde{C}_y$ : Total cross-flow force fluctuation coefficient, $\tilde{C}_y^P$ : Pressure component, $\tilde{C}_y^V$ : Viscous component. The instantaneous power ($\tilde{C}_{yv}, \tilde{C}_{yv}^P, \tilde{C}_{yv}^V$) due to the corresponding force component is also shown (blue line, left axis)). [ $m^* = 2.0, \text{Re}_D = 150, U_r = 6.0, \zeta = 0, L = 20D$ ]

Further, the instantaneous energy transfer between the fluid and the cylinder can be quantified using the temporal variation of the fluid-force coefficients in phase with the cylinder velocity for one complete oscillation cycle[46]. Figure 8 depicts these components ($\tilde{C}_{yv}, \tilde{C}^P_{yv}, \tilde{C}^V_{yv}$) (in blue line on left axis), which are defined as:

$$\tilde{C}_{yv} = \frac{\sqrt{2}\,\tilde{C}_y \cdot v}{\sqrt{v^2}}, \quad \tilde{C}^P_{yv} = \frac{\sqrt{2}\,\tilde{C}^P_y \cdot v}{\sqrt{v^2}}, \text{ and } \tilde{C}^V_{yv} = \frac{\sqrt{2}\,\tilde{C}^V_y \cdot v}{\sqrt{v^2}} \qquad (6)$$

The sing of these components determines if the energy is transferred from the fluid to the cylinder or vice versa. Negative values of these components represent damping offered by the flow against the motion of the cylinder, while positive values indicate that the fluid supports the motion of the cylinder. We observe that the time-averaged net total energy transfer for all the cases is almost zero. For the case of TR = $\infty$ and 40, the pressure component of the fluid force in phase with the cylinder velocity ($\tilde{C}^P_{yv}$) is mostly positive (supporting the oscillation), while the viscous component ($\tilde{C}^V_{yv}$) is primarily negative (presenting damping to the oscillation) over the complete oscillation cycle. Whenever the pressure component dampens the oscillation (i.e. negative $\tilde{C}^P_{yv}$ values), the viscous component supplies the required power to oscillate (i.e. positive $\tilde{C}^V_{yv}$ values), and vice versa. This is different for the cylinder with high taper (i.e. TR = 20 and 12). Being in phase with each other, the pressure component and viscous component contribute similarly (while the viscous component's contribution is relatively low compared to the pressure component). This results in high aerodynamic lift values over highly tapered cylinders.

## B. VIV Characteristics of Tapered cylinders

The previous sections discuss the flow characteristics associated with various taper ratios, focusing on the response observed during the lock-in at a constant reduced velocity of 6. It may show variation in the behavior based on the change in the reduced velocity. Hence, to cover the whole VIV spectrum at $\text{Re}_D = 150$, we have looked at the VIV response of the cylinders with different taper ratios for a wide range of reduced velocities ($3 \leq U_r \leq 16$) in the present section. This section is divided into two subsections, and the first presents the oscillation amplitude and frequency response along with the aerodynamic coefficients (rms of the lift and the mean drag coefficient calculated based on the average cylinder diameter). In the second subsection, to understand the force and oscillation behavior, we have decomposed the force coefficient into two components, in phase with the velocity and the acceleration, respectively. This component of the force coefficient in phase with velocity represents the force exerted on the cylinder due to the relative motion between the fluid and the cylinder. In other words, it captures the effect of the fluid velocity on the forces acting on the cylinder. The component of force coefficient in phase with acceleration represents the force exerted on the cylinder due to its acceleration or deceleration. It accounts for the inertial effects (such as added mass, fluid inertia, etc.) of cylinder motion changes.

## 1. Effect of Reduced Velocities

The 1$^{st}$ row of Fig. 9 shows the variation of the oscillation amplitude over the wide range of the reduced velocities, which covers the whole VIV spectrum. The nature of the VIV for all taper ratios is self-limiting, where the oscillation amplitude initially increases with the increase in the reduced velocity. Eventually, after a particular reduced velocity, it starts to drop with the increasing reduced velocity. It becomes almost constant at a much lower value for the rest of the values of the reduced velocity. All the cylinders show a two-branch response (IB [shaded in yellow in Fig. 9] and LB [shaded in green in Fig. 9]), although the transition from one branch to another occurs at different reduced velocities. The transition from IB to LB leads to a sudden drop in the lift coefficient of the cylinder. It occurs at $U_r = 4$ for TR = $\infty$, 40, 20, and at $U_r = 5$ for TR = 12 (see 3$^{rd}$ row of Fig. 9). Also, the LB branch ends at different reduced velocities for different taper ratios (i.e. $4 \leq U_r \leq 8$ for TR = $\infty$, $4 \leq U_r \leq 9$ for TR = 40, $4 \leq U_r \leq 11$ for TR = 20, and $5 \leq U_r \leq 12$ for TR = 12). This shows that the range of the lock-in varies with the taper ratio. The most extensive lock-in range is observed for TR = 12, and the maximum oscillation amplitude occurs at the much later reduced velocity ($U_r = 9$). The broader spectrum of the high-amplitude oscillations makes it more favorable for energy harvesting applications. The peak amplitude for the other cases (TR = $\infty$, 40, and 20) occurs at the reduced velocities of 4, 6, and 8, respectively (see 1$^{st}$ row of Fig. 9). Thus, tapering the cylinder results in the shift of the peak of the max oscillation amplitude or in-turn the shift in the transitioning of the response branches.

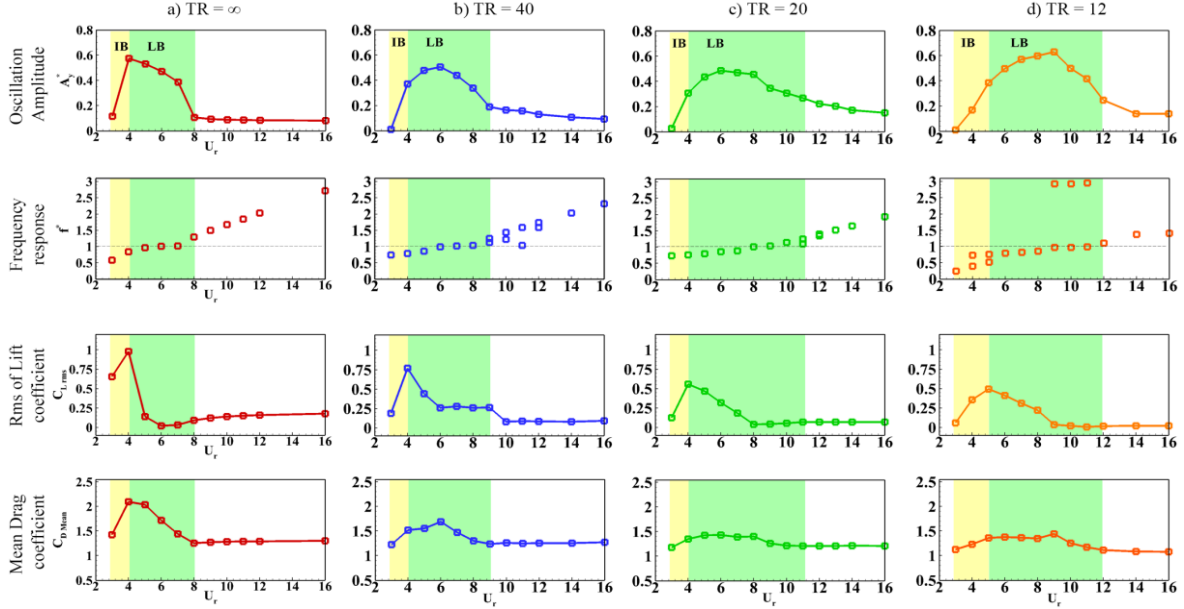

**FIG. 9.** Effect of reduced velocity in terms of the oscillation amplitude ($A_y^*$) [1st row], the frequency response ($f^*$) [2nd row], rms of lift coefficient ($C_{Lrms}$) [3rd row], and the mean drag coefficient ($C_{Dmean}$) [4th row] for different taper ratios: (a) TR = $\infty$, (b) TR = 40, (c) TR = 20, and (d) TR = 12. [$m^* = 2.0, \text{Re}_D = 150, \zeta = 0, L = 20D$]

Further, the 2nd row of Fig. 9 records the dominant frequencies present in the oscillation response of the cylinder for various reduced velocities. We observe the qualitative similarity of the tapered cylinder's frequency response with the uniform cylinder (TR = $\infty$). The lock-in occurs at the natural frequency of the cylinder for the uniform cylinder case, while it occurs at a lower frequency for the case of the tapered cylinders. The frequency response outside the lock-in regime follows the Strouhal law (i.e. the oscillation frequency outside the lock-in is equal to the Strouhal frequency obtained from the stationary cylinder case of the respective taper ratio). The cylinder oscillates with the multi-frequencies in the frequency response in Fig. 9. The lift coefficient increases with the reduced velocities (in IB). It shows a sudden drop as it transitions to LB for all the cases. The last row of Fig. 9 reports the streamwise force coefficient (i.e. mean drag coefficient) for the range of reduced velocities. We observe a minimal change in the values of the mean drag coefficient over the whole spectrum of reduced velocities for the tapered cylinders compared to the uniform cylinder (where the drag coefficient seems to be following the pattern of lift coefficient or the oscillation amplitude). The peak of the mean drag coefficient is observed to be at the same reduced velocity as the peak oscillation amplitude. The following subsection investigates this force and amplitude relation in detail by decomposing the forces into multiple components.

## 2. Force Decomposition, Energy Transfer and Phase Dynamics

To further establish the understanding of the relationship between the amplitude response and the corresponding transverse force, we have plotted the fluctuating amplitude of the lift coefficient ( $C_y = \sqrt{2}(F_y - \tilde{F}_y)_{rms}/(0.5\rho D L U_\infty^2)$ ) against the reduced velocities in the top row of Fig. 10. This force coefficient is further decomposed into two components: in phase with the acceleration: $C_{y,a}$, and in phase with the cylinder's velocity: $C_{y,u}$. Following the available literature[46,47], we have calculated the decomposed components as the product between the instantaneous transverse force fluctuations about its mean and the displacement and velocity of the cylinder, respectively:

$$C'_{y,a} = \frac{\sqrt{2}(F_y - \tilde{F}_y)}{0.5\rho D L U_\infty^2} \frac{y}{\sqrt{\overline{y^2}}} \quad (7)$$

$$C'_{y,u} = \frac{\sqrt{2}(F_y - \tilde{F}_y)}{0.5\rho D L U_\infty^2} \frac{v}{\sqrt{\overline{v^2}}} \quad (8)$$

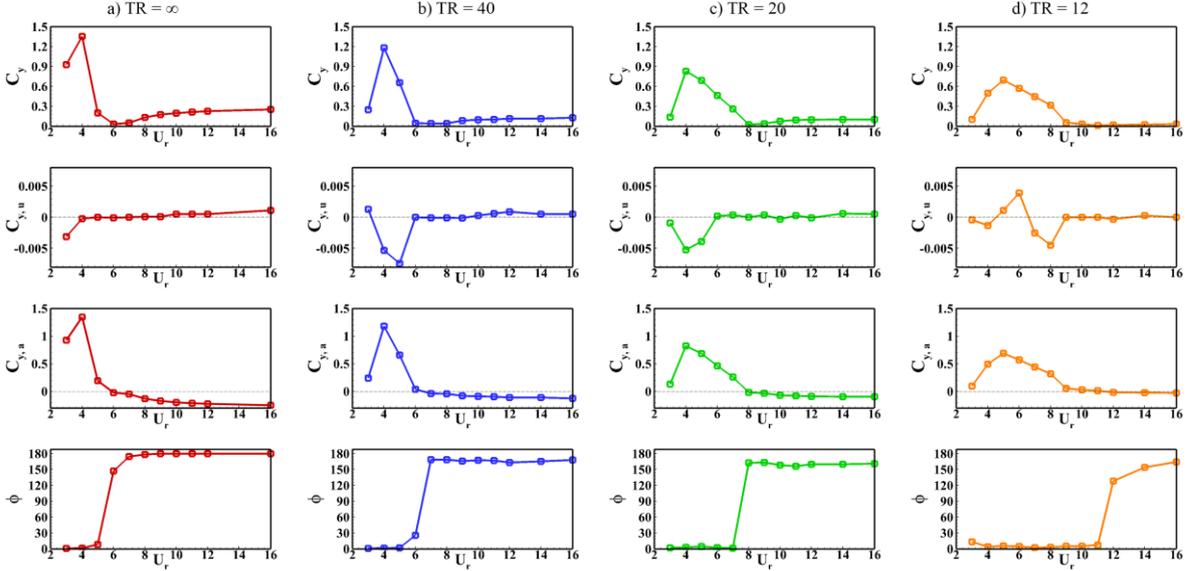

**FIG. 10.** Variation of the fluctuating part of the lift-coefficient ( $C_y$ ) [1st row] and its corresponding component in phase with velocity ( $C_{y,u}$ ) [2nd row], and in phase with acceleration ( $C_{y,a}$ ) [3rd row] along with the phase difference between the lift and the transverse oscillation amplitude ( $\phi$ ) [4th row] at different reduced velocities ( $U_r$ ) for different taper ratios: (a) TR = $\infty$, (b) TR = 40, (c) TR = 20, and (d) TR = 12. [ $m^* = 2.0, \text{Re}_D = 150, \zeta = 0, L = 20D$ ]

The top row of Fig. 10 shows the transverse force variation with the reduced velocities increase, which is already discussed in the preceding subsection and added here to better compare its decomposed components. The 2nd and 3rd row of Fig. 10 reports the variation of the time-averaged values of these coefficients ( $C_{y,u} = \overline{C'_{y,u}}$ and $C_{y,a} = \overline{C'_{y,a}}$) with the reduced velocities ($3 \leq U_r \leq 16$). The component of the lift coefficient in phase with the acceleration ($C_{y,a}$) is related to the fluid inertia, while the component in phase with the cylinder's velocity ($C_{y,u}$) denotes the damping offered by the fluid such that it can be used to quantify the energy transferred from the fluid to the cylinder.

The transverse force coefficient increases with the increase in the reduced velocities in the initial branch until the IB to LB transition occurs. The peak of this coefficient is observed to be at the transition $U_r$ for all taper ratios (refer to the top row of Fig. 9). From the observation of the force component in phase with the velocity ($C_{y,u}$), the energy transfer from the fluid to the tapered cylinder decreases in the initial branch. Negative values of this component mean that the flow oscillations in the transverse direction are not aligned with the motion of the cylinder, and the energy is transferred from the cylinder to the fluid. This implies destructive interference between the motion of the cylinder and the transverse flow, potentially leading to decreased energy in the wake. In most cases, it exhibits negative values during the IB branch. After the transition from IB to LB, the negative coefficient values decrease to change their sign, promoting the energy transfer from the fluid to the cylinder. It becomes zero at the higher reduced velocities, representing the less energetic VIV response. Similarly, the component in phase with the acceleration ($C_{y,a}$) follows a trend similar to the transverse force coefficient. It attains the peak value at the transition $U_r$ for the case of a tapered cylinder. The positive values of this component signify a resonance condition where the fluid forces reinforce the cylinder's motion, potentially leading to increased vibration amplitudes. This coefficient also becomes close to zero for the higher reduced velocities, indicating the less effect of added mass.

Further, we have calculated the phase difference between the transverse force ($F_Y$) and the cylinder's oscillations (Y) using Hilbert transform as:

$$\phi = \phi_{F_Y} - \phi_Y \qquad (9)$$

The last row of Fig. 10 depicts the variation of the phase difference with the reduced velocities. The cylinder oscillations remain in phase with the transverse force for the initial branch for all taper ratios (i.e. the value $\phi$ remains close to zero). Just after the transition to the lower branch, the phase difference between the transverse force and the cylinder oscillations attains a value close to 180° and remains at that value for the remaining reduced velocities. For the case of TR = 12, this transition of the phase difference occurs much later at higher reduced velocities. This can be attributed to structures associated with the third fundamental frequency (refer to the second row of Figure 8) after the transition of IB to LB, which keeps the cylinder oscillations in phase for larger values of the reduced velocities.

To summarize, the present study assessed the flow field associated with the taper cylinders and compared it with the fluid dynamic response of the uniform cylinder at a low Re value of 150. Three-dimensional CFD simulations have been performed over three different taper ratios (i.e. TR = 40, 20, and 12) along with the uniform cylinder (i.e. TR = $\infty$) to develop a basic understanding of the flow features. We have looked at the fast Fourier transform spectrum, the oscillation amplitude and frequency response, the force decomposition, and the corresponding phase dynamics to better characterize the behavior of the tapered cylinders (with mild, moderate, and high taper). The behavior is studied for the undamped cases over a wide range of reduced velocities for the complete closure of the problem.

## V. CONCLUSION

The present study numerically investigates the flow-induced vibrations of a tapered circular cylinder at a Reynolds number equal to 150, bounded to oscillate in a transverse direction only. First, the three-dimensional CFD simulations have been performed on a fixed reduced velocity to assess the fluid dynamic characteristics associated with the wake flow. Then, the VIV behavior was assessed over a wide range of the reduced velocities for three different taper ratios. The results are compared to the uniform cylinder (with TR = $\infty$). The key findings of the study are summarized as follows:

*Wake And Vortex Dynamics*: The simulations are performed for three different taper ratios (i.e. mild (TR = 40), moderate (TR = 20), and high (TR = 12)), and their wake characteristics are compared at a fixed value of the reduced velocity of 6 (which lies in the lock-in regime for all the cases). Due to the geometric shear (spanwise variation of the local diameter) offered to the incoming flow, the Re varies spanwise. Hence, there exists a difference between the vortex shedding modes spanwise. During the one complete cycle, cylinders with all the taper ratios shed two single vortices resembling the 2S pattern of the vortex shedding. As we increase the taper, the wake behind the cylinder exhibits three-dimensional behavior. The appearance of the three-dimensional structures begins from the top end (i.e. the end with a larger diameter) and is much more evident for the TR = 20 and 12. Although, the TR = 40 exhibits the two-dimensional oblique vortex shedding. The spanwise correlation in the pressure (for TR = 20 and 12) drops over the span (from the bottom end to the top end). It shows the minimum at the spanwise location where the two-dimensional nature of the wake changes to three-dimensional.

*Oscillation amplitude and frequency response:* The tapered cylinder's oscillation amplitude response is self-limiting for the range of the reduced velocities (similar to the uniform cylinder). The circular cylinder for all taper ratios exhibits the two-branch amplitude response (i.e. the initial and lower branches). However, the transition from one branch to another is observed to be shifted to higher reduced velocities with the increase in the taper ratio. Similarly, the range of the lock-in is found to be spanned over the extensive range of the reduced velocities as compared to the uniform cylinder case. The case of high taper (i.e. TR = 12) exhibits the widest lock-in regime, with the presence of the third harmonics and the fundamental natural frequency. The frequency is equal to the natural frequency during the lock-in and varies linearly outside the lock-in regime.

*Force decomposition and phase dynamics:* The transverse force is decomposed into two components: in phase with the velocity and phase with the acceleration. These components exhibit negative values during the lower branch, opposing the cylinder oscillations. The peak of these components is observed to occur at the transition $U_r$ for all taper ratios. The components attain positive values after the IB to LB transition, indicating the energy transfer from the fluid to the cylinder. This results in the oscillation being supported by the fluid. The phase dynamics show that the phase difference between the transverse force and the cylinder oscillations is almost $0°$ in IB, and it attains the value $180°$ after the IB to LB transition, resulting in a gradual decrease in oscillation amplitude with the reduced velocities and finally attaining a very low constant oscillation amplitude value.

## DATA AVAILABILITY

The data that support the findings of this study are available from the corresponding author upon reasonable request.

## ACKNOWLEDGMENTS

The authors would like to acknowledge the National Supercomputing Mission (NSM) for providing the computational resources of 'PARAM Sanganak' at IIT Kanpur, which is implemented by C-DAC and supported by the Ministry of Electronics and Information Technology (MeitY) and Department of Science and Technology (DST), Government of India. The authors would also like to acknowledge the IIT-K Computer center (www.iitk.ac.in/cc) for providing the resources to perform the computation work.